\documentclass[11pt]{article}
\usepackage{geometry} % see geometry.pdf on how to lay out the page. There's lots.
\usepackage[english]{babel}
\usepackage[english]{layout}

\usepackage{mathpazo} % math & rm
\usepackage[scaled]{helvet} % ss
\usepackage{courier} % tt
\normalfont
\usepackage[T1]{fontenc}

\usepackage{array} % for better arrays (eg matrices) in maths
\usepackage{paralist} % very flexible & customisable lists (eg. enumerate/itemize, etc.)
\usepackage{verbatim} % adds environment for commenting out blocks of text & for better verbatim
\usepackage[utf8]{inputenc} % Any characters can be typed directly from the keyboard, eg éçñ
\usepackage{textcomp} % provide lots of new symbols
\usepackage{graphicx}  % Add graphics capabilities
\usepackage{epstopdf} % to include .eps graphics files with pdfLaTeX
\usepackage{flafter}  % Don't place floats before their definition

\usepackage{amsfonts,latexsym,amsthm,amssymb}
\usepackage{amsmath,amssymb,bbm,amsfonts}  % Better maths support & more symbols
\usepackage{bm}  % Define \bm{} to use bold math fonts

\usepackage{memhfixc}  % remove conflict between the memoir class & hyperref
\usepackage{fancyhdr} % This should be set AFTER setting up the page geometry
\usepackage{sectsty}
\allsectionsfont{\sffamily\mdseries\upshape} % (See the fntguide.pdf for font help)
\numberwithin{equation}{section}

\begin{document}

\title{\LARGE \textbf{Trigonometric $s\ell(2)$ Gaudin model with boundary terms}}

\author{\textsf{N. ~Cirilo ~Ant\'onio,}
\thanks{E-mail address: nantonio@math.ist.utl.pt}
\textsf{ ~ N. ~Manojlovi\'c}
\thanks{E-mail address: nmanoj@ualg.pt}
\textsf{ and Z. ~Nagy}
\thanks{E-mail address:
zoltan.nagy@m4x.org} \\
\\
\\
\textit{$^{\ast}$Centro de An\'alise Funcional e Aplica\c{c}\~oes}\\
\textit{Instituto Superior T\'ecnico, Universidade T\'ecnica de Lisboa} \\
\textit{Av. Rovisco Pais, 1049-001 Lisboa, Portugal} \\
\\
\textit{$^{\dag\ddag}$Grupo de F\'{\i}sica Matem\'atica da Universidade de Lisboa} \\
\textit{Av. Prof. Gama Pinto 2, PT-1649-003 Lisboa, Portugal} \\
\\
\textit{$^{\dag}$Departamento de Matem\'atica, F. C. T.,
Universidade do Algarve}\\
\textit{Campus de Gambelas, PT-8005-139 Faro, Portugal}\\
}
\date{}

%\keywords{Lattice systems; Exactly solvable models; Gaudin model}

\maketitle
\thispagestyle{empty}
\vspace{10mm}
\begin{abstract}
We review the derivation of the Gaudin model with integrable boundaries. Starting from the non-symmetric R-matrix of the inhomogeneous spin-\textonehalf \ XXZ chain and generic solutions of the reflection equation and the dual reflection equation, the corresponding Gaudin Hamiltonians with boundary terms are calculated. An alternative derivation based on the so-called classical reflection equation is discussed.

\end{abstract}

\clearpage
\newpage

\section{Introduction}

Gaudin models have applications in many areas of modern physics, from quantum optics \cite{Garraway11,BogoliubovKulish12} to physics of metallic nano-grains, see \cite{AmicoOsterloh12} and reference therein.
A model of interacting spins in a chain was first considered by Gaudin \cite{Gaudin76,Gaudin83}. In his approach, these models were introduced as a quasi-classical limit of the integrable quantum chains. Moreover, the Gaudin models were extended to any simple Lie algebra, with arbitrary irreducible representation at each site of the chain \cite{Gaudin83}.

The rational $s\ell(2)$ invariant model was studied in the framework of the quantum inverse scattering method \cite{Sklyanin89}. The quantum inverse scattering method (QISM) \cite{TakhFadI,SklyTakhFad,KulishSklyanin82,Faddeev} as an approach to construct and solve quantum integrable systems has lead to the theory of quantum groups \cite{Drinfeld, Jimbo85}. In his studies, Sklyanin used the $s\ell(2)$ invariant classical r-matrix \cite{Sklyanin89}. A generalization of these results to all cases when skew-symmetric r-matrix satisfies the classical Yang-Baxter equation \cite{BelavinDrinfeld} was relatively straightforward \cite{SklyaninTakebe,Semenov97}. Therefore, considerable attention has been devoted to Gaudin models corresponding to the the classical r-matrices of simple Lie algebras \cite{Jurco89,Jurco90,WagnerMacfarlane00} and Lie superalgebras \cite{BrzezinskiMacfarlane94,KulishManojlovic01,KulishManojlovic03,LimaUtiel01,KurakLima04}. The spectrum and corresponding eigenfunctions were found using different methods such as coordinate and algebraic Bethe ansatz, separated variables, etc. Correlation functions for Gaudin models were explicitly calculated as combinatorial expressions obtained from the Bethe ansatz \cite{Gaudin83}. In the case of the $s\ell(2)$ Gaudin system, its relation to Knizhnik-Zamolodchikov equation of conformal field theory \cite{BabujianFlume,FeiginFrenkelReshetikhin,ReshetikhinVarchenko} or the method of Gauss factorization \cite{Sklyanin99}, provided alternative approaches to computation of correlation functions. The non-unitary r-matrices and the corresponding Gaudin models have been studied recently, see \cite{Skrypnyk09} and the references therein.

A way to introduce non-periodic boundary conditions compatible with the integrability of the quantum systems solvable by the QISM was developed in \cite{Sklyanin88}. The boundary conditions at the left and right sites of the system are expressed in the left and right reflection matrices. The compatibility condition between the bulk and the boundary of the system takes the form of the so-called reflection equation \cite{Cherednik84,KulishSklyanin92}. The compatibility at the right site of the model is expressed by the dual reflection equation. The matrix form of the exchange relations between the entries of the Sklyanin monodromy matrix are analogous to the reflection equation. Together with the dual reflection equation they yield the commutativity of the open transfer matrix \cite{Sklyanin88,FreidelMaillet91,FreidelMaillet91a,KulishSasaki93}.

The starting point to obtain the Gaudin model in the framework of the QISM is the monodromy matrix of the corresponding inhomogeneous spin chain \cite{Sklyanin89}. Hikami, Kulish and Wadati showed that the quasi-classical expansion of the  transfer matrix of the periodic chain, calculated at the special values of the spectral parameter, yields the Gaudin Hamiltonians \cite{HikamiKulishWadati92,HikamiKulishWadati92a}. Hikami generalized this approach to the case of
non-periodic boundary conditions \cite{Hikami95}. In his work Hikami used the Sklyanin monodromy matrix of the open inhomogeneous spin chain. By the quasi-classical expansion of the open transfer matrix, at the special values of the spectral parameter, the corresponding Gaudin Hamiltonians with boundary terms were obtained \cite{Hikami95}. These results were later extended to non-diagonal reflection matrices \cite{YangZhangGould04,YangZhang12} and generalized for other classes of Gaudin models, like for example elliptic ones \cite{YangZhangSasakic04}, as well as to other simple Lie algebras \cite{YangZhangSasakic05}. Analogous results were obtained for the open Gaudin models based on Lie superalgebras \cite{Lima09}.

As it was noticed above, in the periodic case, the Gaudin models are based on classical r-matrices which have a unitarity property. Namely, the density of the Gaudin Hamiltonians coincides with the classical r-matrices and the condition of their commutativity is nothing else but the classical Yang-Baxter equation \cite{Semenov97,Jurco90}. In the case of non-periodic boundary conditions the derivation of the generating function of the corresponding Gaudin Hamiltonians is still an open problem. Sklyanin has developed an approach based on the classical reflection equation, involving a unitary classical r-matrix and corresponding reflection matrix \cite{Sklyanin86,Sklyanin87}. But the derivation of the corresponding Hamiltonians is not straightforward. A method based on a non-unitary classical r-matrix, obtained from a unitary solution of the classical Yang-Baxter equation and the corresponding reflection matrix, was proposed \cite{Skrypnyk09,Skrypnyk07,Skrypnyk10}. However, the corresponding Hamiltonians do not coincide with the ones obtained by the quasi-classical expansion.

This paper is organized as follows. In Section II the R-matrix of the XXZ model and its properties are reviewed. The equivalence between the reflection equation and the dual reflection equation is shown and the corresponding reflection matrix is given explicitly. In Section III an inhomogeneous XXZ spin chain with $N$ sites is studied. It is shown how the quasi-classical expansion of the transfer matrix, for special values of the spectral parameter, yields the Gaudin Hamiltonians, both in the case of periodic and non-periodic boundary conditions. These Hamiltonians are calculated in Section IV. In particular, the
Gaudin Hamiltonians with the boundary terms are given explicitly. An approach to study open Gaudin model based on classical r-matrix, classical reflection equation and corresponding reflection matrix is discussed in Section V.

\section{Reflection equation}
In the framework of the QISM  \cite{TakhFadI,SklyTakhFad,KulishSklyanin82,Faddeev} the starting point in the study of a quantum solvable system is an R-matrix. Here we consider the R-matrix of the spin-\textonehalf \ XXZ  chain \cite{KulishSklyanin80,Jimbo86,Bazhanov}
\begin{equation}
\label{XXZ-Rmatrix}
R (\lambda, \eta) = \left(\begin{array}{cccc}
1 & 0 & 0 & 0 \\[1ex]
0 & \displaystyle{\frac{\sinh(\lambda)}{\sinh(\lambda - \eta)}} & \displaystyle{\frac{-e^{-\lambda}\sinh(\eta)}{\sinh(\lambda - \eta)}} & 0 \\[1.5ex]
0 & \displaystyle{\frac{-e^{\lambda} \sinh(\eta)}{\sinh(\lambda - \eta)}} & \displaystyle{\frac{\sinh(\lambda)}{\sinh(\lambda - \eta)}}  & 0 \\[1.5ex]
0 & 0 & 0 & 1\end{array}\right),
\end{equation}
here $\lambda$ is the so-called spectral parameter and $\eta$ is the quasi-classical parameter. The R-matrix satisfies the Yang-Baxter equation in the space $\mathbb{C}^2 \otimes \mathbb{C}^2 \otimes \mathbb{C}^2$
\begin{equation}
\label{YBE}
R_{12} ( \lambda - \mu) R_{13} ( \lambda) R_{23} (\mu) = R_{23} (\mu ) R_{13} (\lambda ) R_{12} ( \lambda - \mu),
\end{equation}
we suppress the dependence on the quasi-classical parameter $\eta$ and use the standard notation of the QISM  \cite{TakhFadI,SklyTakhFad,KulishSklyanin82,Faddeev} to denote spaces $V_j, j=1, 2, 3$ on which corresponding $R$-matrices $R_{ij}, ij = 12, 13, 23$ act non-trivially. In the present case $V_1 = V_2 = V_3 =\mathbb{C}^2$.

This form of the  R-matrix is related with the symmetric one $R_{12}^t ( \lambda,\eta) = R_{12} ( \lambda, \eta)$ by the similarity transformation
\begin{equation}
\label{similarity-trans}
R_{12} ( \lambda, \eta) \to \mathrm{Ad} \exp (\lambda S^z _1) R_{12} ( \lambda, \eta) ,
\end{equation}
with $S^z = \mathrm{diag}(1/2,-1/2)$ \cite{MezincescuNepomechie91,MezincescuNepomechie91a, MezincescuNepomechie92,Doikou05,Kulish09}. The transformed R-matrix still obeys the Yang-Baxter equation due to the $U(1)$ symmetry of the initial R-matrix
\begin{equation}
\label{0-weght}
[S^z _1 + S^z _2, R_{12} ( \lambda, \eta) ] = 0.
\end{equation}

Sometime ago it was observed \cite{Sklyanin89} that the Gaudin models are related to the classical r-matrix and therefore it is essential that the R-matrix has the quasi-classical property
\begin{equation}
\label{Rsemiclass}
    R (\lambda, \eta) = \mathbbm{1} + \eta r(\lambda) + \mathcal{O} (\eta ^2),
\end{equation}
here $r(\lambda)$ is the corresponding classical r-matrix
\begin{equation}
\label{r-classical}
r(\lambda) = \frac{1}{\sinh (\lambda)} \left(\begin{array}{cccc}0 & 0 & 0 & 0 \\0 & \cosh (\lambda) & - e^{-\lambda} & 0 \\0 & - e^{\lambda}  & \cosh (\lambda) & 0 \\0 & 0 & 0 & 0\end{array}\right),
\end{equation}
which has the unitarity property 
\begin{equation}
\label{r-unitarity}
r_{21}(-\lambda) = - r_{12}(\lambda) ,
\end{equation}
and satisfies the classical Yang-Baxter equation
\begin{equation}
\label{classicalYBE}
[r_{13} (\lambda), r_{23}(\mu) ] + [r_{12}(\lambda - \mu), r_{13} (\lambda) +  r_{23}(\mu)] =0.
\end{equation}
Moreover, for the purpose of deriving the Gaudin Hamiltonians, it is necessary that
the R-matrix \eqref{XXZ-Rmatrix} is normalized so that \cite{HikamiKulishWadati92, HikamiKulishWadati92a}
\begin{equation}
\label{Rlambdazero}
    R (0, \eta) = \mathcal{P},
\end{equation}
where $\mathcal{P}$ is the permutation matrix in $\mathbb{C}^2 \otimes \mathbb{C}^2$.

The R-matrix \eqref{XXZ-Rmatrix} has the unitarity property
\begin{equation}
\label{unitarity}
R_{12} ( \lambda) R_{21} ( - \lambda) = \mathbbm{1},
\end{equation}
the PT-symmetry
\begin{equation}
\label{PT-symmetry}
R_{12}^t ( \lambda) = R_{21} ( \lambda),
\end{equation}
and the following crossing symmetry property
\begin{equation}
\label{R-crossingXXZ}
R ( \lambda) =  \frac{\sinh(\lambda)}{\sinh(\lambda - \eta)} \mathcal{J} _1 R ^{t_2}( -\lambda + \eta ) {\mathcal{J}}^{-1} _1,
\end{equation}
where $t_2$ denotes the transpose in the second space and the matrix $\mathcal{J}$ is given by
\begin{equation}
\label{Q-mat}
\mathcal{J} =  \left(\begin{array}{cc}0 & e^{-\eta/2} \\- e^{\eta/2} & 0 \end{array}\right).
\end{equation}
We observe that the matrix
\begin{equation}
\label{M-mat}
M = \mathcal{J}^t \mathcal{J} = \left(\begin{array}{cc}e^{\eta} & 0 \\ 0 & e^{-\eta} \end{array}\right)
\end{equation}
commutes with the R-matrix
\begin{equation}
\label{MR-commutator}
[M \otimes M, R (\lambda) ] = 0.
\end{equation}
Notice that the unitarity property \eqref{unitarity} and the crossing symmetry \eqref{R-crossingXXZ} imply the following useful identity \cite{MezincescuNepomechie91,Doikou05}
\begin{equation}
\label{crossing-unitarity}
R_{12}^{t_2}(\lambda) M_1^{-1} R_{12}^{t_1}(-\lambda + 2\eta) M_1= \rho (\lambda - \eta, \eta) \mathbbm{1},
\end{equation}
with
\begin{equation}
\label{func-rho}
\quad  \rho (\lambda, \eta) = \frac{\sinh^2(\lambda)-\sinh^2(\eta)}{\sinh^2(\lambda)}.
\end{equation}

A way to introduce non-periodic boundary conditions which are compatible with the integrability of the bulk model, was developed in \cite{Sklyanin88}. Boundary conditions on the left and right sites of the system are encoded in the left and right reflection matrices $K^-$ and $K^+$. The compatibility condition between the bulk and the boundary of the system takes the form of the so-called reflection equation \cite{Cherednik84,KulishSklyanin92}. It is written in the following form for the left reflection matrix acting on the space $\mathbb{C}^2$ at the first site $K^-(\lambda) \in \mathrm{End} (\mathbb{C}^2)$
\begin{equation}
\label{RE}
R_{12}(\lambda - \mu) K^-_1(\lambda) R_{21}(\lambda + \mu) K^-_2(\mu)=
K^-_2(\mu) R_{12}(\lambda + \mu) K^-_1(\lambda) R_{21}(\lambda - \mu) .
\end{equation}

In complete generality, the compatibility on the right end of the model is encoded in the following dual reflection equation \cite{Sklyanin88,FreidelMaillet91,FreidelMaillet91a,KulishSasaki93,NADR}
\begin{equation}
\label{rightRE}
A_{12}(\lambda - \mu) K^{+\, t}_1(\lambda) B_{12}(\lambda + \mu) K^{+\, t}_{2}(\mu) =
K_2^{+\, t}(\mu) C_{12} (\lambda + \mu) K_1^{+\, t}(\lambda) D_{12}(\lambda - \mu) .
\end{equation}
where the matrices $A,B,C,D$ are obtained from the $R$-matrix of the reflection equation \eqref{RE} in the following way
\begin{align}
A_{12}(\lambda)&=\left(R_{12}(\lambda)^{t_{12}}\right)^{-1}=D_{21}(\lambda) , \\
B_{12}(\lambda)&= \left( \left(R_{21}^{t_1}(\lambda)\right)^{-1}\right)^{t_2}=C_{21}(\lambda) .
\end{align}
However, due to the property \eqref{crossing-unitarity} the dual reflection equation \eqref{rightRE} can be written in the equivalent form
\begin{align}
\label{dRE}
&R_{12}( -\lambda + \mu )K_1^{+}(\lambda) M_2 R_{21}(-\lambda - \mu - 2\eta) M_2^{-1} K_2^{+}(\mu)= \notag\\
&K_2^{+}(\mu) M_1 R_{12}(-\lambda -\mu-2\eta) M_1^{-1} K_1^{+}(\lambda) R_{21}(-\lambda + \mu) .
\end{align}
One can then verify that the mapping
\begin{equation}
\label{bijectionKpl}
K^+(\lambda)= K^{-}(- \lambda -\eta) \ M
\end{equation}
is a bijection between solutions of the reflection equation and the dual reflection equation. After substitution of \eqref{bijectionKpl} into the dual reflection equation \eqref{dRE} and using the symmetry property \eqref{MR-commutator} one gets the reflection equation \eqref{RE} with shifted arguments.

Although the K-matrix for the XXZ model is well known \cite{GhoshalZamolodchikov,InamiKonno,VegaGonzalez}, the solution corresponding to the R-matrix \eqref{XXZ-Rmatrix} differs from that matrix (see \cite{Doikou05,KulishMudrov,KMN3}),
\begin{equation}
\label{K-tilde}
\widetilde{K}^{-} (\lambda) =
\left(\begin{array}{cc}
f + e^{-2\lambda} a & (e^{2\lambda}-e^{-2\lambda}) b \\
(e^{2\lambda}-e^{-2\lambda}) c & f + e^{2\lambda} a
\end{array}\right)
\end{equation}
which we normalize
\begin{equation}
\label{Kminus}
K^{-} (\lambda) = \frac{1}{d (\lambda)} \widetilde{K}^{-} (\lambda),
\end{equation}
where
\begin{equation}
\label{d-function}
d (\lambda) = f + a \cosh (2\lambda) + (\sqrt{a^2 + 4 bc} )\sinh  (2\lambda),
\end{equation}
so that
\begin{equation}
\label{normalizationK}
K^{-} (\lambda) K^{-} (- \lambda) = \mathbbm{1}.
\end{equation}
Finally, we observe that matrix $K^{-} (\lambda)$ does not depend on the quasi-classical parameter $\eta$,
\begin{equation}
\label{etaKminus}
\frac{\partial K^{-} (\lambda)}{\partial \eta} = 0.
\end{equation}
The right reflection matrix $K^+(\lambda)$ is obtained by substituting \eqref{Kminus} into \eqref{bijectionKpl}. It is important to notice that
\begin{equation}
\label{normalizationKpl}
\left( \lim_{\eta \to 0} K^+(\lambda) \right) K^{-} (\lambda) = \mathbbm{1}.
\end{equation}
In general, the normalization conditions \eqref{normalizationK}  and \eqref{normalizationKpl} are not essential in  the study of the open spin chain. However, together with \eqref{Rsemiclass} and \eqref{Rlambdazero} they enable the quasi-classical expansion of the transfer matrix which yields the open Gaudin model.

\section{Inhomogeneous XXZ spin chain}

In this section we study an inhomogeneous XXZ spin chain with $N$ sites, characterized by the local space $V_j= \mathbb{C}^2$ and inhomogeneous parameter $\alpha _j$. For simplicity, we start by considering periodic boundary conditions. The Hilbert space of the system is
$$
\mathcal{H} = \underset {j=1}{\overset {N}{\otimes}}  V_j = (\mathbb{C}^2 ) ^{\otimes N}.
$$
In the QISM \cite{TakhFadI, Faddeev,Kulish09} the so-called monodromy matrix
\begin{equation}
\label{monodromy-T}
T(\lambda ) = R_{0N} ( \lambda - \alpha _N) \cdots R_{01} ( \lambda - \alpha _1)
\end{equation}
is used to describe the system. For simplicity we have omitted the dependence on the quasi-classical parameter $\eta$ and the inhomogeneous parameters $\{ \alpha _j , j = 1 , \ldots , N \}$. Notice that $T(\lambda)$ is a two-by-two matrix in the auxiliary space $V_0 = \mathbb{C}^2$, whose entries are operators acting in $\mathcal{H}$. Due to the Yang-Baxter equation \eqref{YBE}, it is straightforward to check that the monodromy matrix satisfies the RTT-relations \cite{TakhFadI, Faddeev,KulishSklyanin82}
\begin{equation}
\label{RTT}
R_{12} ( \lambda - \mu) \underset {1} {T}(\lambda ) \underset {2} {T}(\mu ) = \underset {2} {T}(\mu )\underset {1} {T}(\lambda ) R_{12} ( \lambda - \mu).
\end{equation}
The above equation is written in the tensor product of the auxiliary space $V_0\otimes V_0 = \mathbb{C}^2 \otimes \mathbb{C}^2$, using the standard notation of the QISM and suppressing the dependence on the quasi-classical parameter $\eta$ and the inhomogeneous parameters $\{ \alpha _j , j = 1 , \ldots , N \}$.

The periodic boundary conditions and the RTT-relations \eqref{RTT} imply that the transfer matrix
\begin{equation}
\label{periodic-t}
t (\lambda ) = \mathrm{tr}_0 T(\lambda) ,
\end{equation}
at different values of the spectral parameter commute,
\begin{equation}
\label{periodic-tt}
[t (\lambda) , t (\mu)] = 0,
\end{equation}
here we have omitted the nonessential arguments.

Following references \cite{HikamiKulishWadati92,HikamiKulishWadati92a} we observe that, due to the normalization of the R-matrix \eqref{Rlambdazero}, the quasi-classical expansion of the transfer matrix, for the special values of the spectral parameter, is given by
\begin{equation}
\label{periodic-Zk}
Z_k = t (\lambda = \alpha _k ) = \mathbbm{1} + \eta H_k + \mathcal{O} (\eta ^2),
\end{equation}
where $H_k$ are the Gaudin Hamiltonians, in the periodic case. The commutativity of the Gaudin Hamiltonians,
\begin{equation}
\label{commHk}
[H_k, H_l] =0,
\end{equation}
is ensured by the commutativity of the transfer matrix \eqref{periodic-tt} and the fact the first term in the above expansion is the identity matrix, due to \eqref{Rsemiclass} and \eqref{Rlambdazero}.

In order to construct integrable spin chains with non-periodic boundary condition, one has to use the Sklyanin formalism \cite{Sklyanin88}. The corresponding monodromy matrix $\mathcal{T}(\lambda)$ consists of the two matrices $T(\lambda)$ \eqref{monodromy-T} and a reflection matrix $K ^{-}(\lambda)$ \eqref{Kminus},
\begin{align}
\label{double-T}
\mathcal{T}(\lambda) &= T(\lambda) K ^{-}(\lambda) T^{-1}(- \lambda)  \notag \\
    &= R_{0N}( \lambda - \alpha _N) \cdots R_{01}(\lambda - \alpha _1) K _0^{-}(\lambda) R_{10}(\lambda + \alpha _1 )
\cdots R_{N0}(\lambda + \alpha _N ),
\end{align}
where, for simplicity, we have suppressed the dependence on the other parameters. By construction, the exchange relations of the monodromy matrix $\mathcal{T}(\lambda)$  in $V_0\otimes V_0$ are
\begin{equation}
\label{exchangeRE}
R_{12}(\lambda - \mu) \underset {1} {\mathcal{T}}(\lambda) R_{21}(\lambda + \mu) \underset {2} {\mathcal{T}}(\mu)=
\underset {2} {\mathcal{T}}(\mu) R_{12}(\lambda + \mu) \underset {1}{\mathcal{T}}(\lambda) R_{21}(\lambda - \mu) ,
\end{equation}
using the notation of \cite{Sklyanin88}. The open chain transfer matrix is given by the trace of $\mathcal{T}(\lambda)$ over the auxiliary space $V_0$ with an extra reflection matrix $K^+(\lambda)$ \cite{Sklyanin88},
\begin{equation}
\label{open-t}
t (\lambda) = \mathrm{tr}_0 \left( K^+(\lambda) \mathcal{T}(\lambda) \right).
\end{equation}
The reflection matrix $K^+(\lambda)$ \eqref{bijectionKpl} is the corresponding solution of the dual reflection equation \eqref{dRE}. The commutativity of the transfer matrix for different values of the spectral parameter
\begin{equation}
\label{open-tt}
[t (\lambda) , t (\mu)] = 0,
\end{equation}
is guaranteed by the dual reflection equation \eqref{dRE} and the exchange relations \eqref{exchangeRE} of the monodromy matrix $\mathcal{T}(\lambda)$.

Analogously to the periodic case, the quasi-classical expansion of the transfer matrix, for special values of the spectral parameter, yields the Gaudin Hamiltonians with the boundary terms \cite{YangZhangGould04}
\begin{equation}
\label{open-Zk}
Z_k = t (\lambda = \alpha _k ) = \mathbbm{1} + \eta H_k + \mathcal{O} (\eta ^2),
\end{equation}
where $H_k$ are the corresponding Gaudin Hamiltonians.  As in the periodic case, the commutativity of the Hamiltonians $H_k$ is guaranteed by the equation \eqref{open-tt} and the normalization conditions \eqref{Rsemiclass}, \eqref{Rlambdazero} and \eqref{normalizationKpl}. The Gaudin Hamiltonians will be calculated in the following section.

\section{Trigonometric $s\ell(2)$ Gaudin Hamiltonians with boundary terms}

In this section we calculate explicitly the Gaudin Hamiltonians.  As shown in \cite{HikamiKulishWadati92, HikamiKulishWadati92a}, in the periodic case, it is straightforward to calculate the first two terms in the expansion \eqref{periodic-Zk}
\begin{align}
\label{firstperiodic-Zk}
Z_k |_{\eta = 0} &= \mathrm{tr}_0 \left( R_{0N} ( \alpha _k - \alpha _N) \cdots R_{0k} ( \alpha _k - \alpha _k) \cdots R_{01} ( \alpha _k - \alpha _1) \right) |_{\eta = 0}  \notag \\
    &= \mathrm{tr}_0 \left( R_{0N} ( \alpha _k - \alpha _N) \cdots \mathcal{P}_{0k} \cdots R_{01} ( \alpha _k - \alpha _1) \right) |_{\eta = 0} \notag \\
    &= \mathrm{tr}_0 \left( \mathcal{P}_{0k} \right) = \mathbbm{1},
\end{align}
and the Gaudin Hamiltonians in the periodic case,
\begin{align}
\label{GH-periodic}
H_k = \frac{\partial Z_k}{\partial \eta} |_{\eta = 0} &= \sum _{l > k }\mathrm{tr}_0 \left( R_{0N} ( \alpha _k - \alpha _N) \cdots \frac{\partial R_{0l}( \alpha _k - \alpha _l)}{\partial \eta} \cdots \mathcal{P}_{0k} \cdots R_{01} ( \alpha _k - \alpha _1) \right) |_{\eta = 0} \notag \\
&+ \sum _{l < k }\mathrm{tr}_0 \left( R_{0N} ( \alpha _k - \alpha _N) \cdots \mathcal{P}_{0k} \cdots \frac{\partial R_{0l}( \alpha _k - \alpha _l)}{\partial \eta} \cdots R_{01} ( \alpha _k - \alpha _1) \right) |_{\eta = 0} \notag \\
 &= \sum _{l > k } \mathrm{tr}_0 \left(  \frac{\partial R_{0l}( \alpha _k - \alpha _l)}{\partial \eta} |_{\eta = 0}\mathcal{P}_{0k} \right) + \sum _{l < k }\mathrm{tr}_0 \left(  \mathcal{P}_{0k} \frac{\partial R_{0l}( \alpha _k - \alpha _l)}{\partial \eta} |_{\eta = 0} \right) \notag \\
 &= \sum _{l \neq k } r_{kl}( \alpha _k - \alpha _l).
\end{align}

In the case of non-periodic boundary, we consider the left and right reflection matrices $K^{-}(\lambda)$ and $K^{+}(\lambda)$  given by \eqref{Kminus} and \eqref{bijectionKpl}, respectively. In order to obtain the expansion \eqref{open-Zk}
in \eqref{open-t} we specify $\lambda = \alpha _k$ and calculate the first term
\begin{align}
\label{firstopen-Zk}
Z_k |_{\eta = 0} &= \mathrm{tr}_0 \left( K^+(\alpha _k) \mathcal{T}(\alpha _k) \right) |_{\eta = 0}
= \mathrm{tr}_0 \left( K_0^+(\alpha _k)  R_{0N} ( \alpha _k - \alpha _N) \cdots \mathcal{P}_{0k} \cdots R_{01} ( \alpha _k - \alpha _1)  \right. \times \notag \\
&\left. \times K _0^{-}(\alpha _k) R_{10}(\alpha _k + \alpha _1 ) \cdots R_{k0}( 2\alpha _k ) \cdots R_{N0}(\alpha _k + \alpha _N ) \right) |_{\eta = 0} \notag \\
&= \mathrm{tr}_0 \left( K_0^-( - \alpha _k) \mathcal{P}_{0k} K _0^{-}(\alpha _k) \right) =  K _k^{-}(\alpha _k)K_k^-( - \alpha _k) =\mathbbm{1}.
\end{align}
In the last step above we have used the normalization \eqref{normalizationK}. The Gaudin Hamiltonians with boundary terms are related to the second term in the expansion \eqref{open-Zk},
\begin{align}
\label{calc-GH-boundary}
H_k = \frac{\partial Z_k}{\partial \eta} |_{\eta = 0} &= \mathrm{tr}_0 \left( \frac{\partial K_0^+(\alpha _k)}{\partial \eta} |_{\eta = 0} \mathcal{P}_{0k}  K _0^{-}(\alpha _k) \right) + \mathrm{tr}_0 \left(  K_0^-( - \alpha _k) \mathcal{P}_{0k} K _0^{-}(\alpha _k)  \frac{\partial R_{k0}( 2 \alpha _k )}{\partial \eta} |_{\eta = 0} \right) \notag \\
&+ \sum _{l > k }\mathrm{tr}_0 \left( K_0^-( - \alpha _k)  \frac{\partial R_{0l}( \alpha _k - \alpha _l)}{\partial \eta} |_{\eta = 0} \mathcal{P}_{0k} K _0^{-}(\alpha _k) \right) \notag \\
&+ \sum _{l < k }\mathrm{tr}_0 \left( K_0^-( - \alpha _k)  \mathcal{P}_{0k}  \frac{\partial R_{0l}( \alpha _k - \alpha _l)}{\partial \eta} |_{\eta = 0} K _0^{-}(\alpha _k) \right) \notag \\
&+ \sum _{l \neq k }\mathrm{tr}_0 \left( K_0^-( - \alpha _k)  \mathcal{P}_{0k} K _0^{-}(\alpha _k) \frac{\partial R_{l0}( \alpha _k + \alpha _l)}{\partial \eta} |_{\eta = 0} \right) .
\end{align}
In the derivation above we have used the fact that the left reflection matrix $ K^{-}( \lambda)$ does not depend  on the quasi-classical parameter $\eta$ \eqref{etaKminus}. Finally, the Gaudin Hamiltonians can be expressed in a more concise form
\begin{equation}
\label{GH-boundary}
H_k = \Gamma_k (\alpha _k ) + \sum _{l \neq k } r_{kl}( \alpha _k - \alpha _l) + \sum _{l \neq k } K _k^{-}(\alpha _k) r_{lk}( \alpha _k + \alpha _l)   K_k^-( - \alpha _k),
\end{equation}
where $r_{ij}(\lambda)$ is the corresponding classical r-matrix \eqref{r-classical} and
\begin{equation}
\label{Gamma-k}
\Gamma_k (\alpha _k ) = K _k^{-}(\alpha _k) \left( \frac{\partial K_k^+(\alpha _k)}{\partial \eta} |_{\eta = 0} + \mathrm{tr}_0 \left(  K_0^-( - \alpha _k) \ \mathcal{P}_{0k} \ r_{{k0}}(2 \alpha _k) \right) \right) .
\end{equation}
Notice that the second term on the righthand side of the \eqref{GH-boundary} coincides with the Gaudin Hamiltonians \eqref{GH-periodic} and that the first and the third term are the boundary terms depending on the reflection matrices $K^-$ and $K^+$. The algebraic Bethe ansatz is used to calculate the spectra of these Hamiltonians \cite{YangZhangGould04,YangZhangSasakic05}.

\section{Classical reflection equation and open Gaudin model}
In this section we discuss an approach to study open Gaudin model based on a classical r-matrix,  classical reflection equation and corresponding reflection matrix. A classical reflection equation can be obtained by substituting \eqref{Rsemiclass} into \eqref{RE}, taking into account \eqref{etaKminus}, and comparing the terms of the first order in $\eta$. However, in the original notation of Sklyanin, using the symmetric r-matrix, the formulas are somewhat more compact \cite{Sklyanin86,Sklyanin87}. To this end we apply the similarity transformation \eqref{similarity-trans} on the classical r-matrix \eqref{r-classical} and obtain the symmetric r-matrix
%, $\mathbb{\mathrm{r}}^t(\lambda)=\mathbb{\mathrm{r}}(\lambda)$,
\begin{equation}
\label{rs-classical}
\mathbf{\mathrm{r}}(\lambda) = \frac{1}{\sinh (\lambda)} \left(\begin{array}{cccc}0 & 0 & 0 & 0 \\0 & \cosh (\lambda) & - 1 & 0 \\0 & - 1  & \cosh (\lambda) & 0 \\0 & 0 & 0 & 0\end{array}\right).
\end{equation}
The unitarity conditions now reads
\begin{equation}
\label{rs-unitarity}
\mathbb{\mathrm{r}}(- \lambda) = - \mathbb{\mathrm{r}}(\lambda).
\end{equation}
This r-matrix satisfies the classical Yang-Baxter equation \eqref{classicalYBE}. Together with the corresponding K-matrix
\begin{equation}
\label{K-classical}
\mathbf{\mathrm{K}} (\lambda) = \frac{1}{d (\lambda)}
\left(\begin{array}{cc}
e^{-\lambda} a + e^{\lambda} f  & (e^{2\lambda}-e^{-2\lambda}) b \\
(e^{2\lambda}-e^{-2\lambda}) c &  e^{\lambda} a + e^{- \lambda} f
\end{array}\right) ,
\end{equation}
the function $d (\lambda)$ given in \eqref{d-function}, it satisfies the classical reflection equation obtained by Sklyanin \cite{Sklyanin86,Sklyanin87}
\begin{equation}
\label{cRE}
\left[\mathbf{\mathrm{r}}_{12} (\lambda-\mu) , \mathbf{\mathrm{K}}_1(\lambda) \mathbf{\mathrm{K}} _2 (\mu) \right]
+ \mathbf{\mathrm{K}}_1(\lambda) \mathbf{\mathrm{r}}_{12} (\lambda + \mu) \mathbf{\mathrm{K}} _2 (\mu)  - \mathbf{\mathrm{K}} _2 (\mu)\mathbf{\mathrm{r}}_{12} (\lambda + \mu) \mathbf{\mathrm{K}}_1(\lambda) = 0.
\end{equation}

It is straightforward to define the Gaudin model, in the case of periodic boundary conditions, by introducing the corresponding  Lax matrix, 
\begin{equation}
\label{L-matrix}
L (\lambda) = \sum _{k=1}^N \mathbf{\mathrm{r}}_{0k}(\lambda - \alpha _{k}).
\end{equation}
The Lax matrix obeys the Sklyanin linear bracket
\begin{equation}
\label{Sklybracket}
\left[ \underset {1}{L} (\lambda) , \underset {2} {L} (\mu) \right] = - \left[ \mathbf{\mathrm{r}} _{12} (\lambda - \mu) , \ \underset {1}{L} (\lambda) + \underset {2} {L} (\mu)\right].
\end{equation}
A direct consequence of the Sklyanin linear bracket and the periodic boundary conditions is the fact that the operator
\begin{equation}
\label{gen-function}
\mathbf{\mathrm{t}} (\lambda) = \mathrm{tr}  L ^2(\lambda)
\end{equation}
commutes for different values of the spectral parameter
\begin{equation}
\label{t-commutes}
\mathbf{\mathrm{t}} (\lambda)  \mathbf{\mathrm{t}} (\mu)  =  \mathbf{\mathrm{t}} (\mu)   \mathbf{\mathrm{t}}  (\lambda),
\end{equation}
and is therefore the generating function of integrals of motion. The residues of $\mathbf{\mathrm{t}} (\lambda)$ at poles $\lambda = \alpha _k$ are the corresponding Gaudin Hamiltonians \cite{Sklyanin89,Jurco89,KulishManojlovic03}.  

A method to study open Gaudin model based on the following Lax matrix
\begin{equation}
\label{Skrypnyk-L}
\mathcal{L} (\lambda) = L (\lambda) - \mathbf{\mathrm{K}} (\lambda) \ L (- \lambda) \ \mathbf{\mathrm{K}} (- \lambda),
\end{equation}
and the corresponding r-matrix
\begin{equation}
\label{Skrypnyk-r}
\mathbf{\mathrm{r^K}} _{12} (\lambda,  \mu)  = \mathbf{\mathrm{r}} _{12} (\lambda - \mu) -{\mathbf{\mathrm{K}}}_{2}  (\mu) \ \mathbf{\mathrm{r}}_{12} (\lambda + \mu) \ {\mathbf{\mathrm{K}}}_{2}  (- \mu),
\end{equation}
was developed recently \cite{Skrypnyk09,Skrypnyk07,Skrypnyk10}. The relevant bracket in this case is given by
\begin{equation}
\label{Skrypnyk-bracket}
\left[ \underset {1}{\mathcal{L}} (\lambda) , \underset {2} {\mathcal{L}} (\mu) \right] = - \left(  \left[ \mathbf{\mathrm{r^K}} _{12} (\lambda,  \mu) , \underset {1}{\mathcal{L}} (\lambda) \right] - \left[ \mathbf{\mathrm{r^K}} _{21} (\mu, \lambda) , \underset {2} {\mathcal{L}} (\mu) \right] \right).
\end{equation}
The operator 
\begin{equation}
\label{Skrypnyk-t}
\mathfrak{t} (\lambda) = \mathrm{tr} \mathcal{L}^2 (\lambda) 
\end{equation}
is considered to be the generating function of integrals of motion \cite{Skrypnyk09,Skrypnyk07,Skrypnyk10}. However, the Hamiltonians obtained as the residues of $\mathfrak{t} (\lambda)$ at the poles $\lambda = \alpha _k$ do not coincide with the Hamiltonians obtained in the previous section. 

\section{Conclusions}

We have reviewed the Gaudin model with integrable boundaries starting from the non-symmetric R-matrix of the spin-\textonehalf \ XXZ chain and generic solutions of the reflection equation and the dual reflection equation. We have shown how the quasi-classical expansion of the transfer matrix of the open inhomogeneous XXZ chain, calculated for special values of the spectral parameter, yields the Gaudin Hamiltonians with the boundary terms. Finally, an alternative approach to study open Gaudin model based on the classical Yang-Baxter equation and classical reflection equation is discussed.

\bigskip

\textbf{Acknowledgments.} This work was supported by the FCT Project No. \hfil \break PTDC/MAT/099880/2008 through the European program COMPETE/FEDER.

\end{document}